\documentclass[twocolumn]{aastex631}

\usepackage{multirow}
\usepackage{amsmath}

\begin{document}

\title{Relations of rotation and chromospheric activity to stellar age for FGK dwarfs from Kepler and LAMOST}

\author[0009-0009-1338-1045]{Lifei Ye}
\affiliation{Institute for Frontiers in Astronomy and Astrophysics, Beijing Normal University, Beijing 102206, China}
\affiliation{Department of Astronomy, Beijing Normal University, Beijing 100875, People’s Republic of China}

\author[0000-0002-7642-7583]{Shaolan Bi}
\affiliation{Institute for Frontiers in Astronomy and Astrophysics, Beijing Normal University, Beijing 102206, China}
\affiliation{Department of Astronomy, Beijing Normal University, Beijing 100875, People’s Republic of China}
\email{bisl@bnu.edu.cn}

\author[0000-0002-2510-6931]{Jinghua Zhang}
\affiliation{South-Western Institute for Astronomy Research, Yunnan University, Chenggong District, Kunming 650500, China}
\email{zhjh@nao.cas.cn}

\author[0000-0003-0795-4854]{Tiancheng Sun}
\affiliation{Institute for Frontiers in Astronomy and Astrophysics, Beijing Normal University, Beijing 102206, China}
\affiliation{Department of Astronomy, Beijing Normal University, Beijing 100875, People’s Republic of China}

\author[0000-0003-2908-1492]{Liu Long}
\affiliation{Institute for Frontiers in Astronomy and Astrophysics, Beijing Normal University, Beijing 102206, China}
\affiliation{Department of Astronomy, Beijing Normal University, Beijing 100875, People’s Republic of China}

\author[0000-0002-2614-5959]{Zhishuai Ge}
\affiliation{Beijing Planetarium, Beijing Academy of Science and Technology, Beijing, 100044, China}

\author[0000-0001-6396-2563]{Tanda Li}
\affiliation{Institute for Frontiers in Astronomy and Astrophysics, Beijing Normal University, Beijing 102206, China}

\author[0000-0002-3672-2166]{Xianfei zhang}
\affiliation{Institute for Frontiers in Astronomy and Astrophysics, Beijing Normal University, Beijing 102206, China}
\affiliation{Department of Astronomy, Beijing Normal University, Beijing 100875, People’s Republic of China}

\author[0000-0003-3957-9067]{Xunzhou Chen}
\affiliation{Research Center for Intelligent Computing Platforms, Zhejiang Laboratory, Hangzhou 311100, China}

\author[0000-0003-3020-4437]{Yaguang Li}
\affiliation{Institute for Astronomy, University of Hawai`i, 2680 Woodlawn Drive, Honolulu, HI 96822, USA}

\author[0009-0004-9024-9666]{Jianzhao Zhou}
\affiliation{Institute for Frontiers in Astronomy and Astrophysics, Beijing Normal University, Beijing 102206, China}
\affiliation{Department of Astronomy, Beijing Normal University, Beijing 100875, People’s Republic of China}

\author{Maosheng Xiang}
\affiliation{CAS Key Laboratory of Optical Astronomy, National Astronomical Observatories, Chinese Academy of Sciences, Beijing 100101, China}
\affiliation{Institute for Frontiers in Astronomy and Astrophysics, Beijing Normal University, Beijing 102206, China}

\begin{abstract}

The empirical relations between rotation period, chromospheric activity, and age can be used to estimate stellar age. To calibrate these relations, we present a catalog, including the masses and ages of 52,321 FGK dwarfs, 47,489 chromospheric activity index $logR^{+}_{HK}$, 6,077 rotation period $P_{rot}$ and variability amplitude $S_{ph}$, based on data from LAMOST DR7, Kepler and Gaia DR3. We find a pronounced correlation among $P_{rot}$, age, and [Fe/H] throughout the main-sequence phase for F dwarfs. However, the decrease of $logR^{+}_{HK}$ over time is not significant except for those with [Fe/H] $<$ $-$0.1. For G dwarfs, both $P_{rot}$ and $logR^{+}_{HK}$ are reliable age probes in the ranges $\sim$ 2-11 Gyr and $\sim$ 2-13 Gyr, respectively. K dwarfs exhibit a prominent decrease in $logR^{+}_{HK}$ within the age range of $\sim$ 3-13 Gyr when the relation of $P_{rot}-\tau$ is invalid. These relations are very important for promptly estimating the age of a vast number of stars, thus serving as a powerful tool in advancing the fields of exoplanet properties, stellar evolution, and Galactic-archaeology.

\end{abstract}

\keywords{\href{http://astrothesaurus.org/uat/1629}{Stellar rotation (1629)}; \href{http://astrothesaurus.org/uat/1580}{Stellar activity (1580)}; \href{http://astrothesaurus.org/uat/230}{Stellar chromospheres (230)}; \href{http://astrothesaurus.org/uat/1581}{Stellar ages (1581)}}

\section{Introduction} \label{sec:intro}

Stellar age plays a crucial role in the field of planetary systems, stellar physics, Galaxy formation and evolution. The age determinations of most dwarfs are difficult because traditional model-dependent methods use stellar properties, which either change little as the star evolves or are hard to observe \citep{stellar_Age_2010ARA&A..48..581S}. The rotation rates and activity levels of all dwarfs significantly decrease over time as these stars gradually lose their angular momentum driven by magneto stellar winds \citep[e.g.,][]{skumanich_law, Noyes1984ApJ...279..763N, Kawaler_AMloss1988ApJ, Matt2005ApJ...632L.135M, meibom_NGC6811, AM_LOSS_2023ApJ...948L...6M}. When appropriately calibrated, rotation period and activity proxy serve as reliable indicators of stellar ages, utilizing the methods of gyrochronology and magnetochronology. Gyrochronology and Magnetochronology are defined as relations among rotation period, activity level, mass, and age \citep[e.g.,][]{barnes2003ApJ...586..464B, barnes2007ApJ...669.1167B, mamajek_rhk'_2008ApJ...687.1264M, vidotto_magnetochronology2014}.

Recently, the gyrochronology and magnetochronology using open clusters have been finely calibrated within 4 Gyr \citep[e.g.,][]{barnes2010ApJ...722..222B, gyro_Garcia_2014A&A...572A..34G, matt_gyro_2015ApJ...799L..23M, MHD2016ApJ...832..145R, zhangjiajun2019, Angus_oc_gyro_2019AJ....158..173A, Curtis_oc_gyro_2019ApJ...879...49C, reprucht147_2020ApJ...904..140C, magnetic_Age_2021QSRv..27407259P, M67_2022ApJ...938..118D, longliu_oc_2023arXiv230706596L}. Although the age range of near-solar metallicity dwarfs with asteroseismic parameters is expanded to $\sim$ 10 Gyr \citep{gyro_aster_Angus_2015MNRAS.450.1787A, VanSaders2016Natur.529..181V, aster_activity_2020MNRAS.491..455B}, the small sample size of data leads to oversimplified formulas that fail to account for the impact of mass and metallicity. In order to calibrate gyrochronology and magnetochronology effectively, it is important to establish the intricate relations among rotation period, magnetic activity, mass, and age for different spectral types. This is necessary that a large sample of accurate measurements of the rotation period and magnetic activity proxies, accomplishing with reliable ages.

With the dramatically improved astrometric data from Gaia DR3 \citep{gaia2022}, we can now achieve high-precision luminosity. Millions of the spectra from LAMOST \citep{LAMOST_2012RAA....12..723Z} not only provide us with homogenous atmospheric parameters ($T_{\rm eff}$, [Fe/H]), but also are used to measure chromospheric activity levels. Combining these observations we can obtain precise mass and age through grid-based modeling \citep{GBM_1993A&AS...98..523S, YY2001ApJS..136..417Y, GBM_2008A&A...486..951G, DESP_2008ApJS..178...89D, Parsec_2012MNRAS.427..127B, MIST_2016ApJ...823..102C, GBM_2023A&A...671A..78M}. In addition, Kepler space mission provides the rotation period ($P_\mathrm{rot}$) and photometric variability amplitudes ($S_\mathrm{ph}$) for tens of thousands of stars \citep{Basri_var_ampli_2011AJ....141...20B, McQ_2013MNRAS.432.1203M, McQ2014prot_ApJS..211...24M, Mathur_Sph_2014A&A...562A.124M, reinhold_rotaion_2015A&A...583A..65R, santos2019, santos2021}. Thanks to these improvements, it will provide a chance to construct a large, homogeneous, and comprehensive sample, which opens a new insight into gyrochronology and magnetochronology.

\begin{figure}[ht!]
\centering
\includegraphics[width=0.48\textwidth]{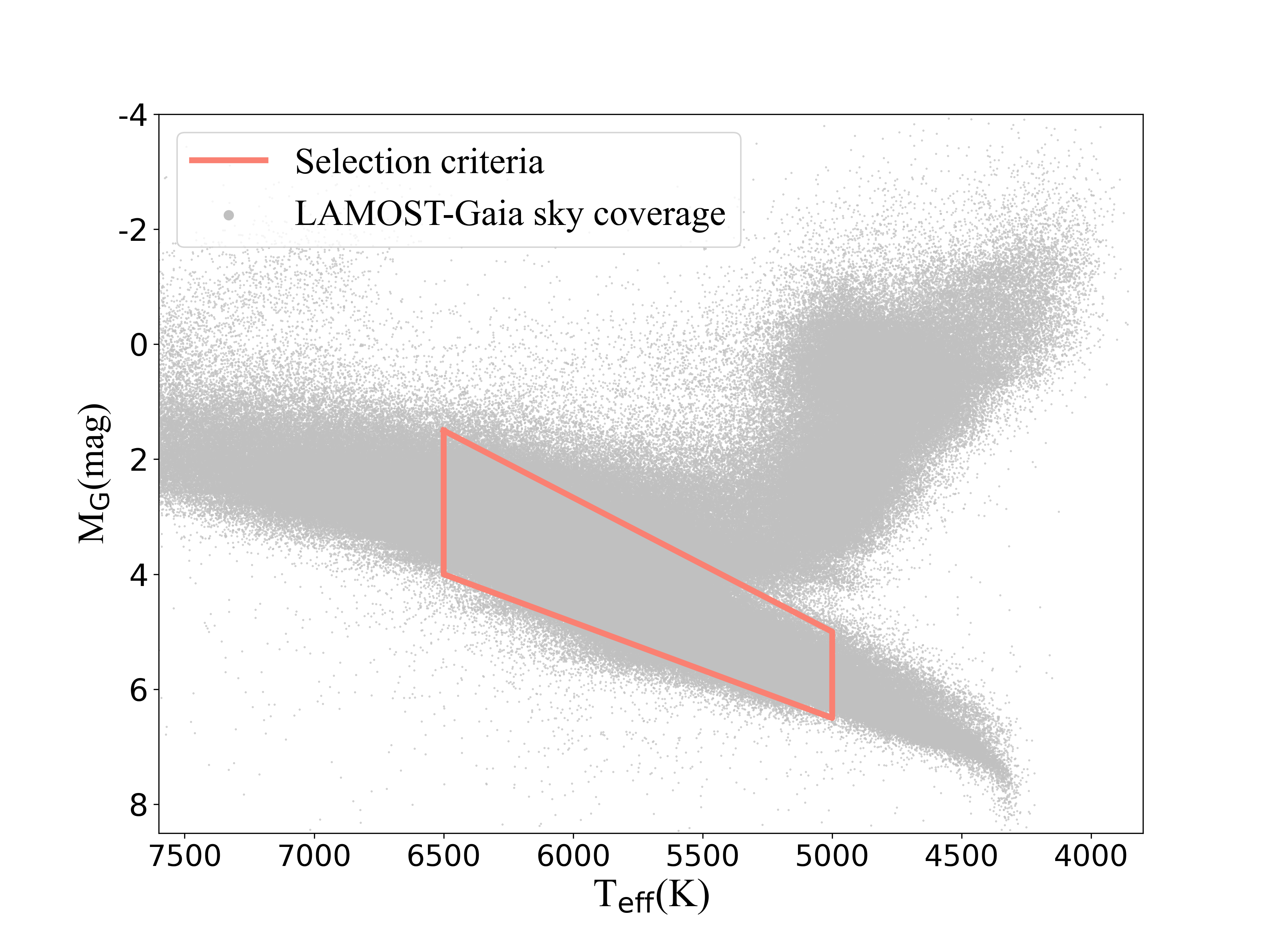}
\caption{$T_{\rm eff}-M_G$ diagram of the stars in the field of LAMOST and Gaia (grey dots); the orange boundary is our selection criteria according to Section.\ref{sec: sample}.
\label{fig: HRD}}
\end{figure}

This work aims to provide a catalog and explore the relations among stellar rotation, magnetic activity, mass and age. The sample selection details are in Section \ref{sec: sample}. The stellar mass and age are estimated in section \ref{sec: estimate}. In Section \ref{sec: result}, we calibrate the relations among the above parameters. Our conclusions are summarized in Section.\ref{sec: summary}.

\section{Sample selection} \label{sec: sample}

We use atmospheric parameters of LAMOST DR7 derived by Data-Driven Paynes (DD-Payne) approach \citep{xiang2019abundance}. The absolute magnitude($M_{G}$) and luminosity are from Gaia DR3\footnote{\url{https://gea.esac.esa.int/archive/documentation/GDR3/Data_analysis/chap_cu8par/sec_cu8par_apsis/ssec_cu8par_apsis_flame.html}}. The sample are selected as the following steps:

\begin{enumerate}
\item[1.] We set $S/N \geq 50$, $flag_{chi2}$\footnote{Reduced $\chi^2$ of the spectral fit.} = 1, and $flag_{unique}\footnote{Flag indicating the selection of the spectrum with the highest S/N during repeated visits.}$ = 1 to obtain reliable atmospheric parameters from LAMOST DR7 data.
\item[2.] Cross-matching stars between Gaia DR3 and LAMOST DR7. Based on the crossover radius of 2 arcsecs, we obtain 1,274,433 common stars with luminosity uncertainty less than 10\%, as shown as gray dots in Fig.\ref{fig: HRD}.
\item[3.] We adopt the following empirical formula to select the main-sequence stars in the $T_{\rm eff}$ range of 5000–6500 K by referring to \cite{Select_sample_2022AJ....163..179P}:
\begin{equation}
    M_G = -0.00233T_{\rm eff} + 16.66667 , \label{eq: upper}
\end{equation}
\begin{equation}
    M_G = -0.00167T_{\rm eff} + 14.83333 . \label{eq: lower}
\end{equation}

the Eq.\ref{eq: upper} and Eq.\ref{eq: lower} correspond to the upper and lower boundaries, respectively. A total of 849,163 main-sequence stars are selected as shown in Fig.\ref{fig: HRD};
\item[4.] We use Gaia renormalized unit weight error (RUWE) $\geq 1.2$ as the criterion to remove non-single stars or astrometric noise \citep{rybizki2022classifier}. We also remove the non-rotation type of the International Variable Star Index\footnote{\url{https://www.aavso.org/vsx/}} (VSX) to reduce binaries or multiple systems.
\end{enumerate}

Based on the above steps, we obtain 698,885 stars with precise $T_{\rm eff}$, [Fe/H], [$\alpha$/Fe] and luminosity, which is defined as the total sample (hereafter TSample). By cross-matching the Tsample with the catalog of \cite{santos2019,santos2021} and removing the classical pulsator and close-in binary candidates (CP/CB), we obtain 7400 stars, including the rotation period ($P_{rot}$), variability amplitude ($S_{ph}$) and their corresponding errors named Sample I. 

The $R^{+}_{HK}$ of dwarfs, derived based on the S-index, is an adequate proxy for the intensity of chromospheric magnetic activity. The single observation for the measurement of the S-index is probably overestimated or underestimated in a long magnetic activity cycle. Therefore, we only adopt the stars with two or more observations from the TSample and then calculate the weighted average S-index by the following formula:
\begin{equation}\label{eq:shk}
    {S}=\frac{\sum^n S_{L}*SNR}{\sum^n SNR},
\end{equation}
where $S_{L}$ is the S-index and SNR is signal to noise ratio, adopted by \cite{zhangweitao2022}. Following the method of \cite{mittag2013} to obtain $R^{+}_{HK}$ of stars. The specific formula is:
\begin{equation}
    R^{+}_{HK}=\frac{\mathcal{F}_{HK}-\mathcal{F}_{HK,phot}-\mathcal{F}_{HK,basal}}{\sigma\mathcal{T}^4_{eff}},
\end{equation}
where $\sigma$ is the Stefan-Boltzmann constant, $\mathcal{F}_{HK, phot}$ is the photometric flux in Ca II H\&K lines, and $\mathcal{F}_{HK, basal}$ is basal chromospheric flux. $\mathcal{F}_{HK}$ is the arbitrary surface flux, defined by \cite{middelkoop1982}: 
\begin{equation}
    {\mathcal{F}_{HK}}=10^{8.25-1.67(B-V)}S,
\end{equation}
\noindent where the B-V values are derived by the interpolation of $T_{\rm eff}$ and [Fe/H] \citep{B-V_inter_2005ApJ...626..465R}.

We obtain $R^{+}_{HK}$ of 61,701 stars, regarded as Sample II. In addition, to compare the chromospheric activity level of the Sun with G dwarfs, we converted solar S-index (0.179 - 0.194) on LAMOST scale \citep{zhang2020} to solar $R^{+}_{HK}$ $\sim (1.209 - 1.600)*10^{-5}$ according to the above process. All the samples used in our work are listed in Table \ref{tab: sample}.

\begin{table}
    \footnotesize
    \centering
    \caption{The summary of samples used in our work. The $\checkmark$ indicates that all the sample stars have corresponding parameters.}
    \label{tab: sample}
    \begin{tabular}{cccc}
        \hline
        \textbf{Parameter} & \textbf{TSample} & \textbf{Sample I} & \textbf{Sample II} \\
        $T_{\rm eff}$, [Fe/H], [$\alpha$/Fe] & \checkmark        & \checkmark  & \checkmark \\
        Luminosity             & \checkmark          & \checkmark  & \checkmark \\
        $P_{rot}$              &                     & \checkmark  &            \\
        $S_{ph}$     &                     & \checkmark  &            \\
        $S$\ index    &                     &             & \checkmark \\
        $R^{+}_{HK}$ &                     &             & \checkmark \\
        Volume                 & 698885\ stars        & 7400\ stars  & 61701\ stars \\
        \hline
    \end{tabular}
\end{table}

\begin{figure*}[ht!]
\centering
\includegraphics[width=0.9\textwidth]{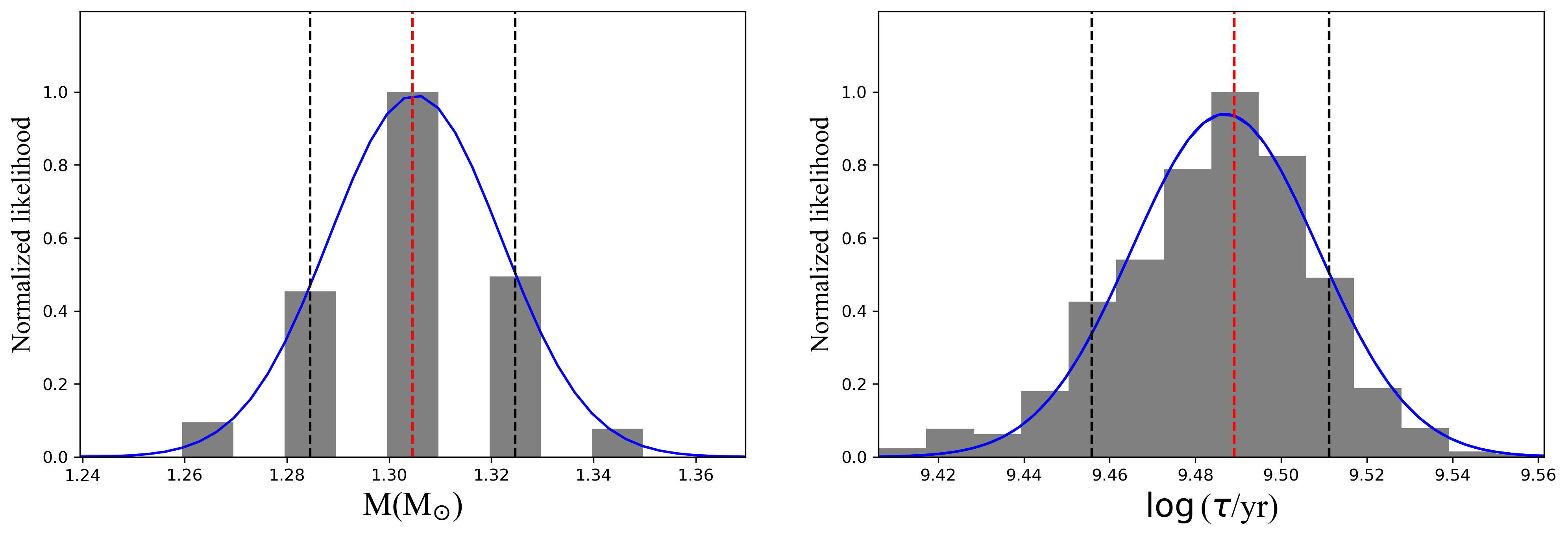}
\caption{Likelihood distributions of mass and age in logarithmic scale for the example star, which is the common target of our sample and asteroseismic sample. The vertical red solid line represents the most probable value, while the black dashed lines indicate the probability values at the 0.16 and 0.84 quantities.}
\label{fig:gaussfit}
\end{figure*}

\section{Fundamental Parameter Estimation} \label{sec: estimate}
\subsection{Stellar Model Grid}\label{subsec:GBM}
Stellar model grid is a method for determining the fundamental characteristics of stars by generating a series of theoretical tracks with different metallicity, masses, and ages, and comparing them to observations such as $T_{\rm eff}$, log\textit{g}, [Fe/H] to determine the best-fit model. The $\alpha$-enhanced stellar evolution model for this work is established in \cite{stc_2023arXiv230704086S}, computed using the Modules for Experiments in Stellar Astrophysics (MESA) code \citep[]{paxton2011modules, paxton2013modules, paxton2015modules, paxton2018modules, paxton2019modules}{}. The inlist file (for MESA) utilized in the computation of our stellar models is available on Zenodo at doi:\dataset[10.5281/zenodo.7866625]{https://zenodo.org/records/7866625}.

The mass range of this stellar model grid is from 0.7 M$_{\odot}$ to 1.5 M$_{\odot}$ with a grid step of 0.02 M$_{\odot}$. The input [Fe/H] values vary from $-$2.00 to +0.45 dex with a grid step of 0.05 dex. Additionally, the $\alpha$-enhanced value ranges from 0.0 to 0.3 dex with a step size of 0.1 dex. The helium enrichment law was calibrated with initial abundances of helium and heavy elements of the standard solar model provided by \cite{paxton2011modules}, resulting in a helium-to-metal enrichment ratio of $Y = 0.248 + 1.3324Z$. The mixing-length parameter $\alpha_{\rm MLT}$ is set to 1.82. In order to account for the effect of microscopic diffusion and gravitational settling of elements in low-mass stars, we employed the formulation of \cite{thoul1994element}, which can modify the surface abundances and main-sequence lifetimes \citep{MS_time_2001ApJ...562..521C, MS_time10.1111/j.1365-2966.2012.21948.x}. We utilized the solar mixture GS98 \citep{GS98_1998SSRv...85..161G} and supplemented the opacity tables with OPAL high-temperature opacities\footnote{\url{http://opalopacity.llnl.gov/new.html}} and low-temperature opacities \citep{ferguson2005}. Other details about the input physics of our stellar models are presented in \cite{stc_2023arXiv230704086S}.

\subsection{Bayesian framework}\label{subsec:fittingfr}

We utilize Bayesian statistics\citep{basu2010} to infer stellar fundamental parameters by combining the prior knowledge of the model-given parameters with observed data. This is expressed as the posterior probability of a model $\textit{M}_{i}$ given data $\mathcal{D}$ and prior information $\mathcal{I}$. The result is the posterior probability distribution of the model stellar parameters, $P(M_{i}|D, I)$.
\begin{equation} \label{overallp}
    P(M_i|D,I) = \frac{p(M_i|I)p(D|M_i)}{p(D|I)}
\end{equation}
where $P(M_{i}|I)$ is the normalized prior probability of a particular model, which is equal to the reciprocal of the number of models $1/N_{m}$. Then:
\begin{equation}\label{likelihood}
\begin{split}
P(D|M_i,I) &= L(T_{\rm eff},[Fe/H],Luminosity)\\
&= L_{T_{\rm eff}}L_{[Fe/H]}L_{Luminosity}
\end{split}
\end{equation}
where L is the maximum likelihood function, in the form of:
\begin{equation}
    L=\frac{1}{\sqrt{2\pi}\sigma} \exp\frac{-(\phi_{obs}-\phi_{model})^2}{2\sigma^2},
\end{equation}
here $\sigma$ is the error of the observation $\phi_{obs}$.

The p(D$|$I) in Eq.\ref{overallp} is a normalization factor for the specific model probability, 
\begin{equation}
    p(D|I) = \sum\limits^{N_m}_{j=1}p(M_j|I)p(D|M_j,I)\label{PDI}
\end{equation}

The simplified version of the posterior probability is obtained by canceling the uniform priors:
\begin{equation}
    p(M_i|D,I) = \frac{p(D|M_i,I)}{\sum^{N_m}_{j=1}p(D|M_j,I)}\label{MDI}.
\end{equation}

\begin{figure}[ht!]
\centering
\includegraphics[scale=0.45]{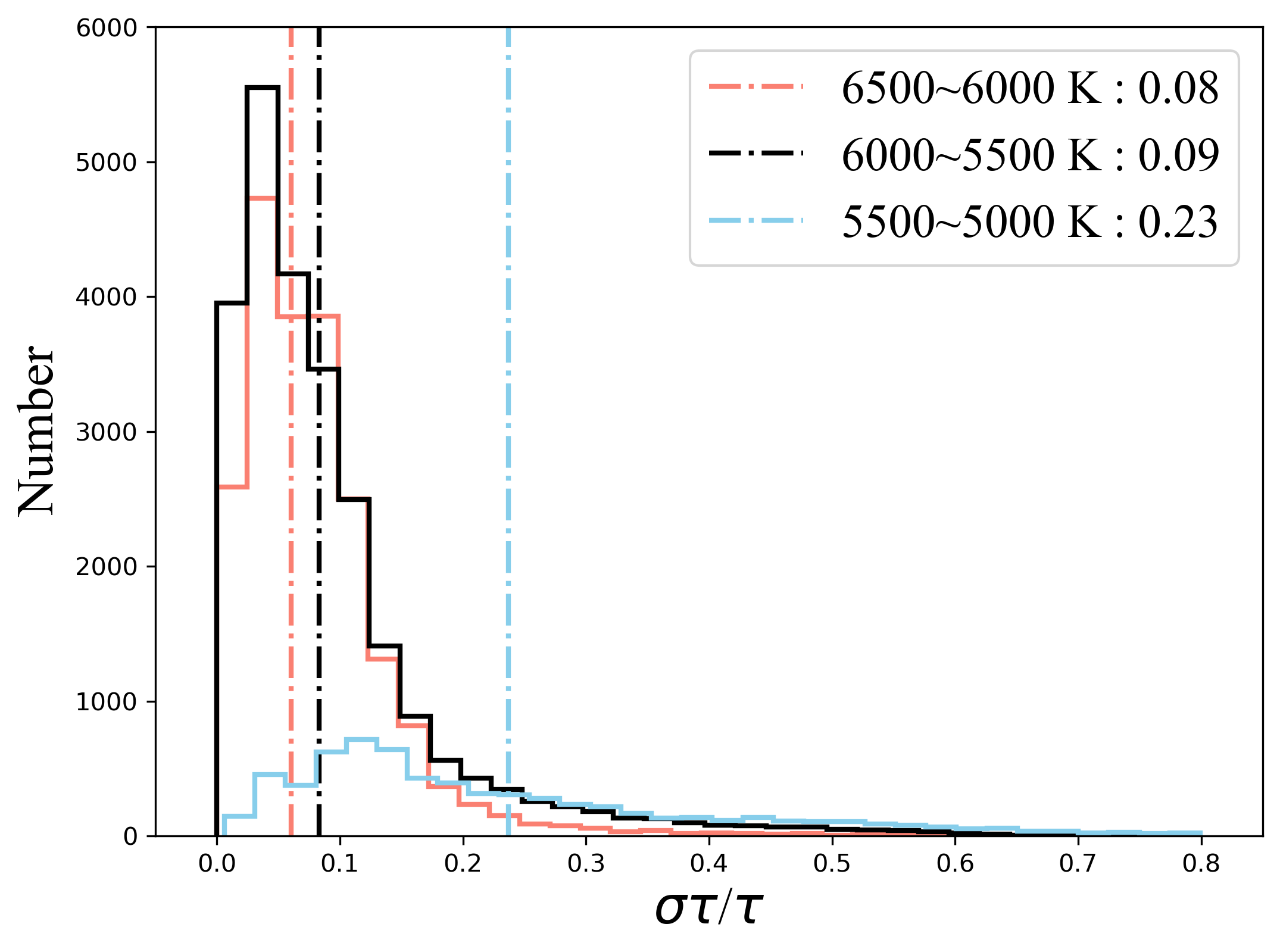}
\caption{Age uncertainty of stars in different effective temperature intervals as labeled in legend.}
\label{fig:uncer_Age}
\end{figure}

The result provides the probability distribution for the selected models with the most likely fundamental parameters, and is plotted in Fig.\ref{fig:gaussfit}. Posterior probability distributions are constructed for each parameter to obtain estimation at the median, along with uncertainty expressed as standard deviation.

To evaluate the reliability of masses and ages in this study, we divide the sample into three temperature ranges of 500 K intervals. The results show that the mean of age precision for the 6500-6000 K, 6000-5500 K, and 5500-5000 K are 8\%, 9\%, and 23\%, respectively, as shown in Fig.\ref{fig:uncer_Age}. A total of 52,321 stars with reliable age uncertainty ($\sigma_{\tau}/\tau \leq 0.8$) and $\tau \leq 13.8$ Gyr are preserved, and the mean age uncertainty is 11\%. 

\begin{figure*}[ht!]
\centering
\includegraphics[width = 0.9\textwidth]{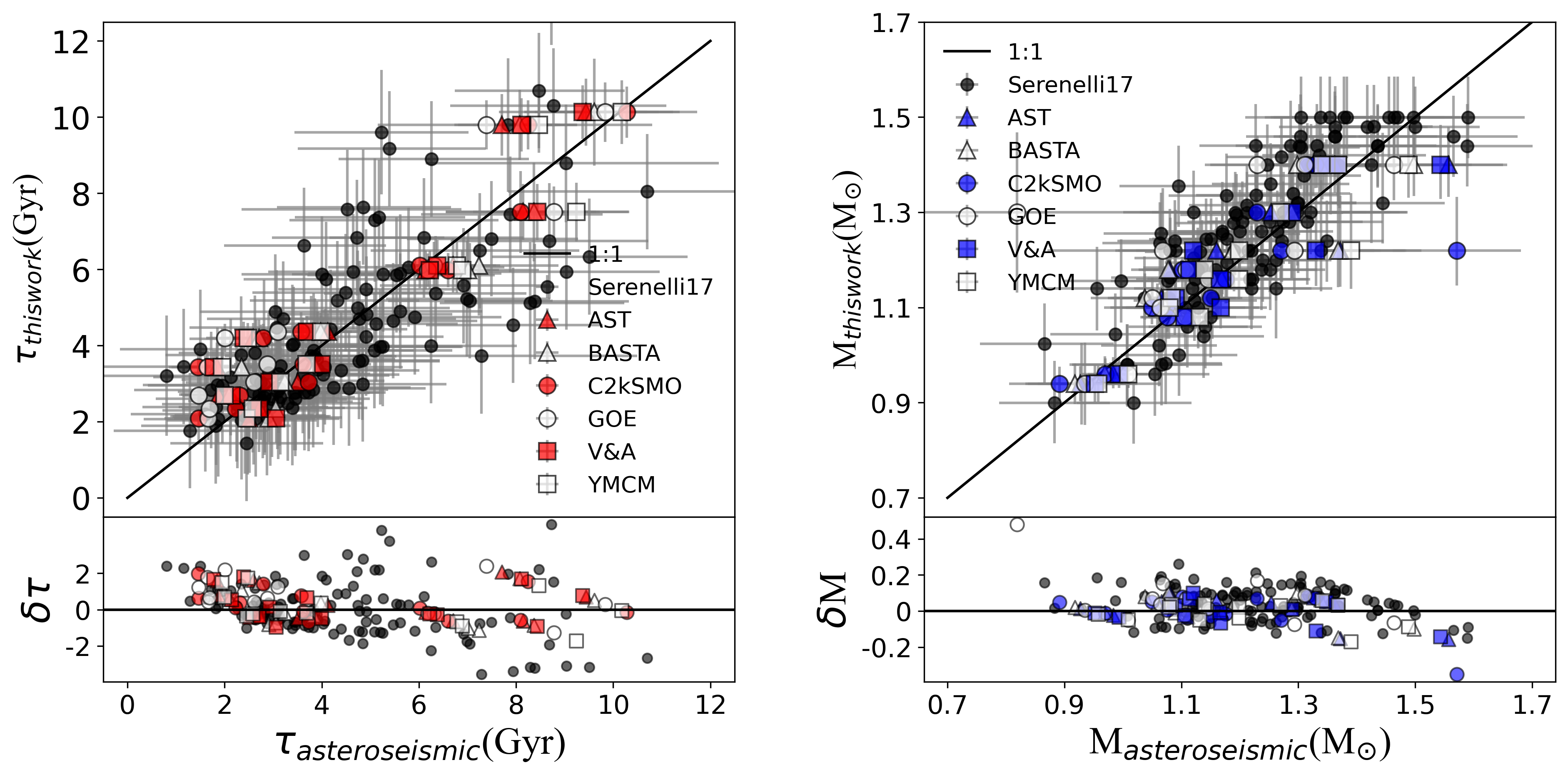}
\caption{Comparison of fundamental parameters between this work and asteroseismic sample. The x-axis displays the mass and age values calculated using asteroseismic parameters. The solid lines represent the equal values of the parameters. Different symbols are used for different pipelines: solid triangle for AST, hollow triangle for BASTA, red spot for C2kSMO, circle for GOE, solid square for V\&A, hollow square for YMCM, and black spot for Serenelli17. The short gray line is the error bar. The residuals are plotted on the Y-axis in the respective lower panels.}
\label{fig:bench_t}
\end{figure*}

In addition, we cross-match our sample with the asteroseismic sample \citep{silva_aster_Age_2017ApJ...835..173S, Serenelli_2017ApJS..233...23S}, and obtain 137 common stars. We compare the masses and ages of these stars derived by seven pipelines with our results, as shown in Fig.\ref{fig:bench_t}. The median absolute deviation (MAD) is listed in Table \ref{tab: aster_mad}. Our results indicate a good agreement with the asteroseismic parameters.

\begin{table}[h!t]
\scriptsize
\centering
\tabcolsep 0.6truecm
\caption{The Median Absolute Deviations (MAD) obtained from comparing our mass and age with those derived from six different pipelines.} \label{tab: aster_mad}
\begin{tabular}{llll}
\hline
\hline
\multicolumn{1}{c}{Pipeline} &
\multicolumn{1}{c}{MAD of age}&
\multicolumn{1}{c}{MAD of mass}
\\
\hline
ASTFIT & 2.52\% & 1.38\% \\
BASTA & 6.06\% & 2.36\% \\
C2kSMO & 5.10\% & 4.16\% \\
GOE & 25.68\% & 3.37\% \\
V\&A & 4.19\% & 0.55\% \\
YMCM & 3.35\% & 1.77\% \\
Serenelli17 & 3.13\% & 3.02\% \\
\hline
\\
\end{tabular}
\tablecomments{The details of the first six pipelines are described in \cite{silva_aster_Age_2017ApJ...835..173S} and the last pipeline in \cite{Serenelli_2017ApJS..233...23S}.}
\end{table}

\section{Results} \label{sec: result}

\begin{figure*}[ht!]
\plotone{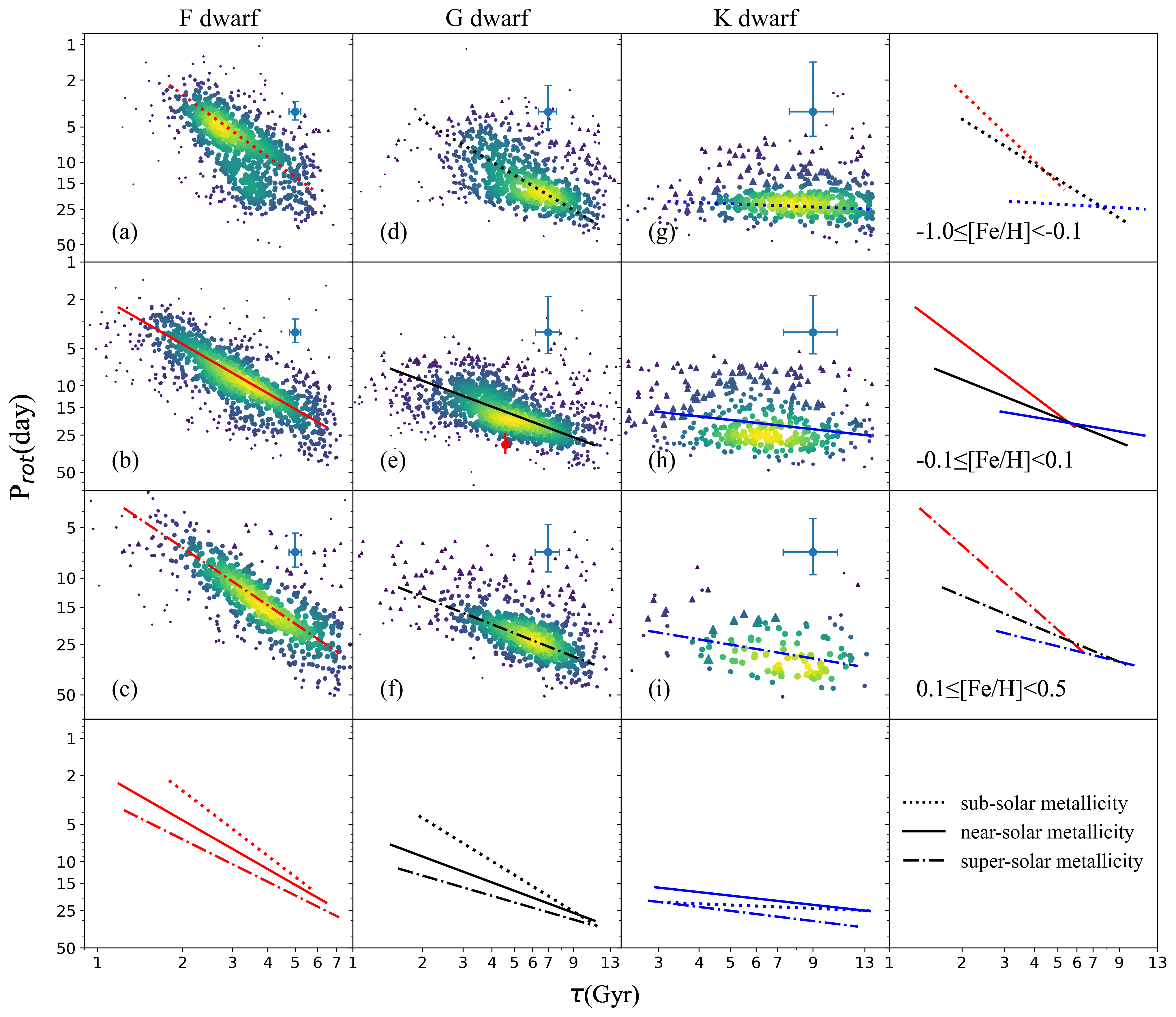}
\caption{Distribution of rotation period with age for different spectral types, as indicated by the title. The segment linear fitting is represented by lines, with line colors corresponding to spectral type and line styles denoting distinct metallicity, as noted in the text. Scatter sizes as well as color reflect the data distribution density. The solar $P_{rot}$ (24.47 $\sim$ 35 day) and age ($\sim$ 4.57Gyr) are marked by a red error bar in the central subplot. Blue error bar is the typical uncertainty of each bin. Triangles represent the $logS_{ph} >$ 3.3 in diffuse points (the data distribution density less than 16\%).}
\label{fig: PTD}
\end{figure*}

\begin{figure*}[ht!]
\plotone{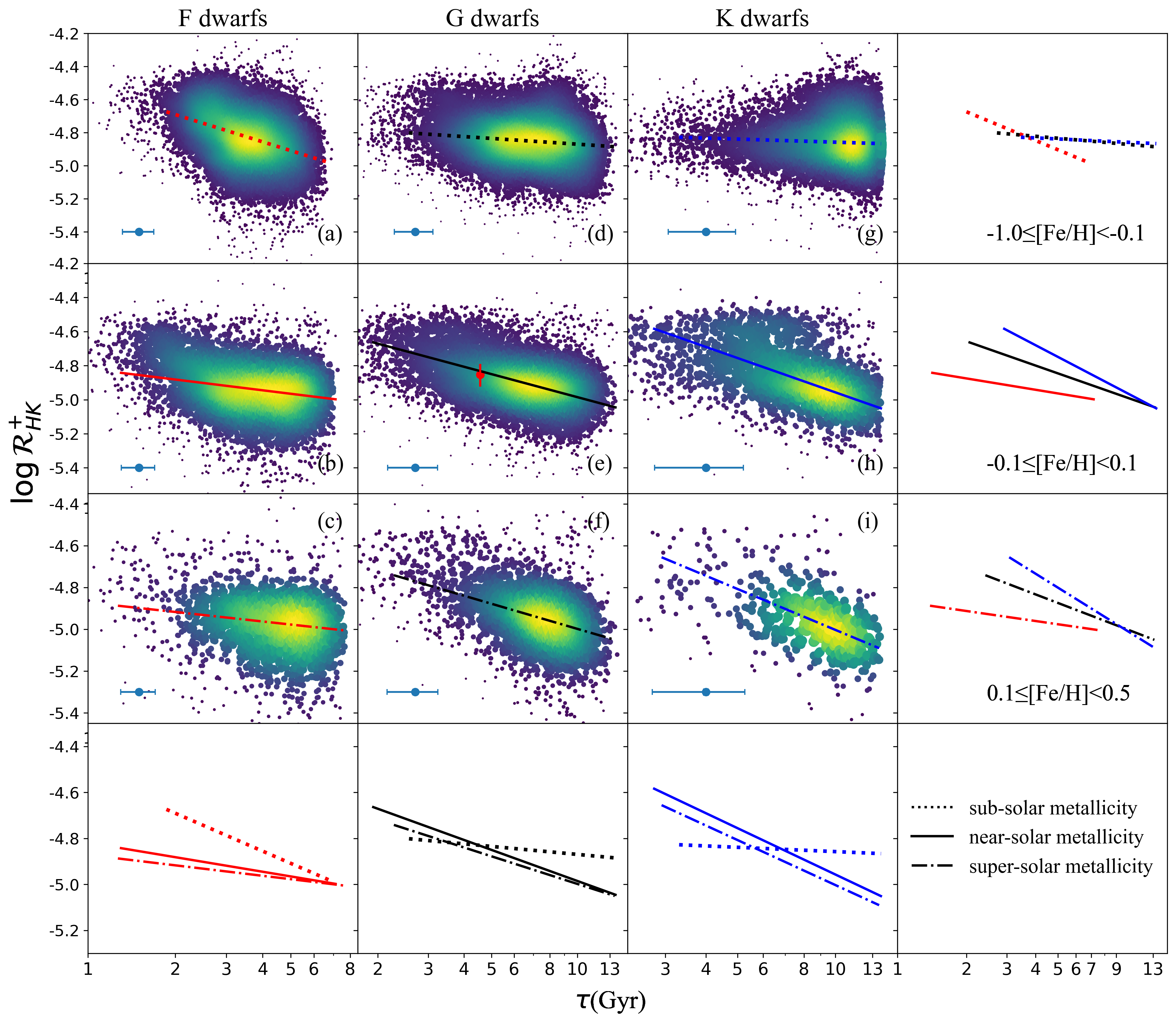}
\caption{Consistent with Fig.\ref{fig: PTD}, except that the y-axis is the chromospheric activity indices, and solar $R^{+}_{HK} \sim (1.209 - 1.600)*10^{-5}$. 
\label{fig: RTD}}
\end{figure*}

\begin{figure*}[ht!]
\plotone{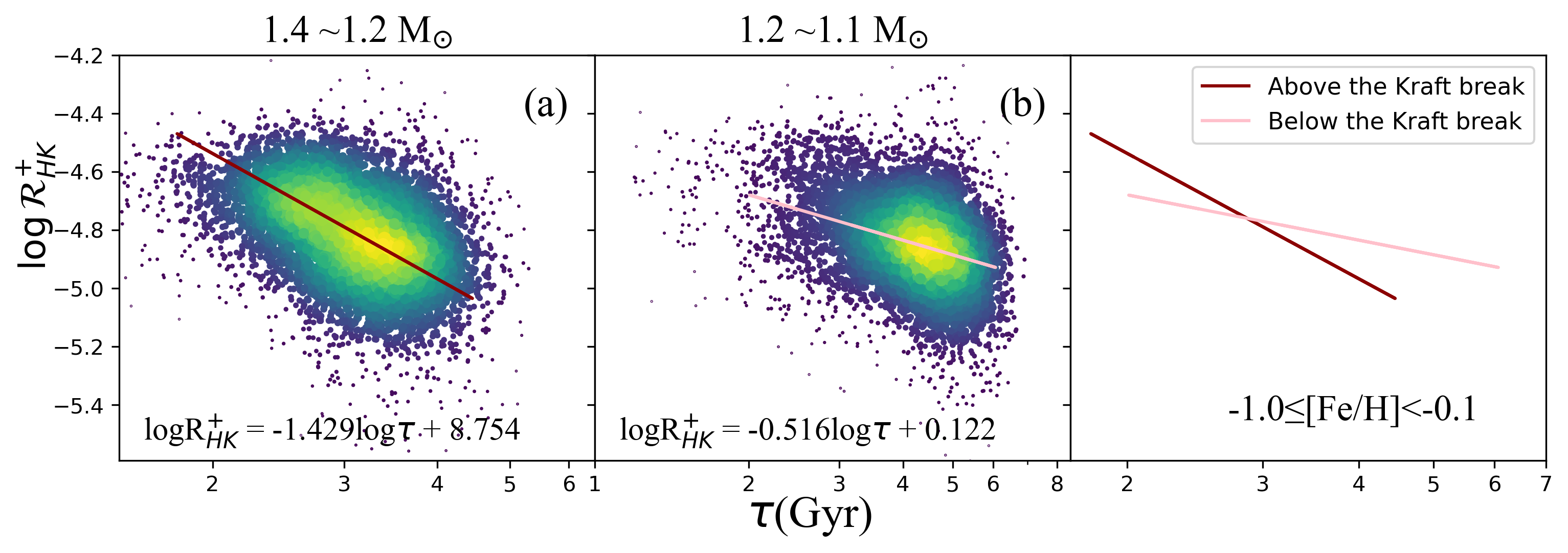}
\caption{The sub-bins of Fig.\ref{fig: RTD}(a), the title indicates the mass range of stars and the fitting results as shown as text and solid line. 
\label{fig: KbRTD}}
\end{figure*}

\begin{table*}[h!t]
\scriptsize
\centering
\tabcolsep 0.17truecm
\caption{Coefficient, intercept of linear fitting in Eq.\ref{linearfitting}. Application range, Pearson correlation coefficient, the null hypothesis for each bin, the number of stars included in the fit, typical uncertainty of x- and y-axis in each panel.} \label{tab: coef_inter}
\begin{tabular}{c c c c c c c c c c c}
\hline
\hline
\multicolumn{1}{l}{y} &
\multicolumn{1}{c}{Mass interval} &
\multicolumn{1}{c}{[Fe/H] interval} &
\multicolumn{1}{c}{$\alpha$} &
\multicolumn{1}{c}{$\beta$}&
\multicolumn{1}{c}{Age range}&
\multicolumn{1}{c}{$r-value$}&
\multicolumn{1}{c}{$p-value$}&
\multicolumn{1}{c}{$N_{fit}$}&
\multicolumn{1}{c}{Age error}&
\multicolumn{1}{c}{y error}
\\
~ & ($M_{\odot}$) & ~ & ~ & ~ & (Gyr) & ~ & ~ & ~ & (Gyr) & ~ \\
\hline
\multirow{9}{*}{$P_{rot}$}&\multirow{3}{*}{1.4 $\sim$ 1.1}&-1.0 $\sim$ -0.1&$1.750\pm0.060$&$-15.850\pm0.567$&1-5&0.597&0.0&848&0.241&0.670\\
~&~&-0.1 $\sim$ 0.1&$1.442\pm0.029$&$-12.782\pm0.274$&1-6&0.816&0.0&1008&0.231&0.800\\
~&~&0.1 $\sim$ 0.5&$1.132\pm0.026$&$-9.707\pm0.246$&1-7&0.868&0.0&583&0.245&1.610\\
\cline{2-11}
~&\multirow{3}{*}{1.1 $\sim$ 0.9}&-1.0 $\sim$ -0.1&$1.159\pm0.042$&$-10.133\pm0.412$&2-11&0.590&0.0&733&0.631&1.480\\
~&~&-0.1 $\sim$ 0.1&$0.731\pm0.024$&$-5.857\pm0.232$&2-11&0.667&0.0&1151&0.835&1.800\\
~&~&0.1 $\sim$ 0.5&$0.546\pm0.024$&$-3.965\pm0.23$&2-11&0.645&0.0&591&0.850&2.210\\
\cline{2-11}
~&\multirow{3}{*}{0.9 $\sim$ 0.7} &-1.0 $\sim$ -0.1&$0.106\pm0.047$&$0.324\pm0.463$&2-13&0.024&0.207&397&1.403&2.290\\
~&~&-0.1 $\sim$ 0.1&$0.285\pm0.086$&$-1.494\pm0.844$&2-13&0.204&0.001&394&1.693&1.835\\
~&~&0.1 $\sim$ 0.5&$0.334\pm0.08$&$-1.839\pm0.792$&2-12&0.364&0.001&105&1.724&2.600\\
\hline
\multirow{9}{*}{$R^{+}_{HK}$}&\multirow{3}{*}{1.4 $\sim$ 1.1} &-1.0 $\sim$ -0.1&$-0.548\pm0.011$&$0.404\pm0.106$&1-6&-0.438&0.0&9954&0.186&0.012\\
~&~&-0.1 $\sim$ 0.1&$-0.208\pm0.010$&$-2.950\pm0.099$&1-7&-0.328&0.0&4960&0.197&0.012\\
~&~&0.1 $\sim$ 0.5&$-0.149\pm0.020$&$-3.532\pm0.193$&1-7&-0.207&0.0&1508&0.204&0.012\\
\cline{2-11}
~&\multirow{3}{*}{1.1 $\sim$ 0.9} &-1.0 $\sim$ -0.1&$-0.119\pm0.007$&$-3.679\pm0.070$&2-13&-0.147&0.0&11632&0.416&0.012\\
~&~&-0.1 $\sim$ 0.1&$-0.449\pm0.009$&$-0.495\pm0.084$&2-13&-0.397&0.0&6220&0.536&0.012\\
~&~&0.1 $\sim$ 0.5&$-0.400\pm0.018$&$-0.997\pm0.182$&2-13&-0.413&0.0&2257&0.544&0.013\\
\cline{2-11}
~&\multirow{3}{*}{0.9 $\sim$ 0.7} &-1.0 $\sim$ -0.1&$-0.061\pm0.017$&$-4.244\pm0.166$&3-13&0.015&0.365&6576&0.936&0.013\\
~&~&-0.1 $\sim$ 0.1&$-0.671\pm0.018$&$1.753\pm0.182$&2-13&-0.616&0.0&2026&1.216&0.013\\
~&~&0.1 $\sim$ 0.5&$-0.651\pm0.042$&$1.506\pm0.416$&2-13&-0.451&0.0&439&1.263&0.015\\
\hline
\\
\end{tabular}
\end{table*}

We present a catalog of 52,321 stars with $P_{rot}$, $S_{ph}$, $R^{+}_{HK}$, [Fe/H], mass and age, listed in Table \ref{tab: col_over}. To limit the relations among rotation, activity, metallicity, mass and age, we divide the sample into 9 bins according to their [Fe/H] and mass. Metallicity bins are classified as:
\begin{itemize}
    \item sub-solar metallicity: $-1.0 \leq [Fe/H] < -0.1$,
    \item near-solar metallicity: $-0.1 \leq [Fe/H] <  0.1$,
    \item super-solar metallicity: $ 0.1 \leq [Fe/H] <  0.5$.
\end{itemize}

Stars at fixed metallicity bin are classified as:

\begin{itemize}
\item F dwarf: $1.4 - 1.1 M_{\odot}$,
\item G dwarf: $1.1 - 0.9 M_{\odot}$,
\item K dwarf: $0.9 - 0.7 M_{\odot}$. 
\end{itemize}

\begin{table*}[h!t]
\scriptsize
\centering
\tabcolsep 0.2truecm
\caption{Columns overview of the Sample.} \label{tab: col_over}
\begin{tabular}{lllllll}
\hline
\hline
\multicolumn{1}{c}{Parameter} &
\multicolumn{1}{c}{Format} &
\multicolumn{1}{c}{Units}&
\multicolumn{1}{c}{Description}&
\\
\hline
Source & long & -- & Gaia DR3 ID \\
RA & float & deg & Right ascension (J2000) \\
DEC & float & deg & Declination (J2000) \\
SNR\_G & float & -- & Spectral $S/N$ per pixel in SDSS g band \\
Teff & float & K & Effective temperature \\
Teff\_err & float & K & Uncertainty in Teff \\
logg & float & dex & Log of surface gravity \\
logg\_err & float & dex & Uncertainty in logg \\
FeH & float & dex & Iron abundance \\
FeH\_err & float & dex & Uncertainty in FeH \\
Alpha\_Fe & float & dex & Alpha element to iron abundance ratio \\
Alpha\_Fe\_err & float & dex & Uncertainty in Alpha\_Fe \\
Gmag & float & mag & Absolute magnitude in Gaia G band \\
Gmag\_err & float & mag & Standard error of Gmag \\
Lum\_Flame & float & $L_{\odot}$ & Stellar luminosity with FLAME pipeline \\
Lum\_Flame\_err & float & $L_{\odot}$ & Standard error of Lum\_Flame \\
KIC & int & -- & Kepler Input Catalog ID \\
Prot & float & day & Rotation Period \\
E\_Prot & float & day & Uncertainty in Prot \\
Sph & float & ppm & Amplitude brightness variations \\
E\_Sph & float & ppm & Uncertainty in Sph \\
S\_index & float & -- & Weight average value of $S_{L}$ index \\
S\_err & float & -- & Uncertainty in S-index \\
SNR & float & -- & S-index $S/N$ \\
logRHKplus & float & -- & Chromospheric activity index \\
logRHKplus\_err & float & -- & Uncertainty in RHKplus \\
Mass & float & $M_{\odot}$ & Mass with grid-based modelling \\
Mass\_err & float & $M_{\odot}$ & Uncertainty in Mass \\
Age & float & Gyr & Age with grid\-based modelling \\
Age\_err & float & Gyr & Uncertainty in Age \\
\hline
\\
\end{tabular}
\tablecomments{The comparison of Mass and T$_{\rm eff}$ in our sample with those from \cite{santos2019, santos2021} indicate that the MAD is $\sim$ 101 K, with a dispersion $\sim$ 172 K for T$_{\rm eff}$; the MAD is $\sim$ 0.015 M$_{\odot}$, with a dispersion value $\sim$ 0.065 M$_{\odot}$ for Mass.} 
\end{table*}

Fig.\ref{fig: PTD} and Fig.\ref{fig: RTD} show the relations of $P_{rot}$($\tau$, M, [Fe/H]) and $R^{+}_{HK}$($\tau$, M, [Fe/H]), respectively. For each bin, we use the Pearson correlation coefficient (r-value) and the null hypothesis (p-value) to examine the potential connections between two parameters as displayed in Table \ref{tab: coef_inter}. Note that the p-value for all bins is less than 0.05, which means the significant linear relation of $logP_{rot}-\tau$ and $logR^{+}_{HK}-\tau$, except for K dwarfs with sub-solar metallicity greater than 0.05. Therefore we can employ a linear fitting model in the forms of the following:

\begin{equation}
    \log y = \alpha \log \tau + \beta , \label{linearfitting}
\end{equation}
where y corresponds to the $P_{rot}$ or $R^{+}_{HK}$. $\tau$ is age in years. $\alpha$ is the slope, representing the spin-down rate in Fig.\ref{fig: PTD} or the decay rate of $logR^{+}_{HK}$ in Fig.\ref{fig: RTD}. $\beta$ is the intercept, representing the constant for each bin. The best fits are listed in Table \ref{tab: coef_inter}. To ensure the robustness of our fitting results, we weigh our fitting with the data distribution density.

The optimal fit is achieved by minimizing the weighted sum of residuals, r($\theta$).
\begin{equation}
    r(\theta) = \sum_{i=1}^{N}w_i(y_i-f(x_i, \theta))^2, \label{weight}
\end{equation}
Where N is the number of data points. $x_i$ and $y_i$ are the data point. $\theta$ represents the free parameter. $w_i$ is the weight, which is determined by the data distribution density. In this work, we sorted stars by age and computed the data distribution density using the Gaussian kernel density estimation \citep{scott2015multivariate} for individual points within 3 $\sigma$:
\begin{equation}
    \hat{f}(x) = \frac{1}{Nh}\sum_{i=1}^{N}\frac{1}{\sqrt{2\pi}} \exp\frac{-(x-x_i)^2}{2h^2}, \label{PDF}
\end{equation}
where h is the bandwidth that controls the width of the Gaussian kernel.

\subsection{$P_{rot}$($\tau$, M, [Fe/H])}\label{subsec:PT}

According to the correlation test and the slope of linear fitting, we find a pronounced relation among $P_{rot}$, metallicity, mass and age, as shown in Fig.\ref{fig: PTD}. For FG dwarfs in Fig.\ref{fig: PTD}(a-c) and Fig.\ref{fig: PTD}(d-f), the results clearly indicate that: i) $P_{rot}$ increase with increasing age within a range of 1-7 Gyr for F dwarfs, and 2-11 Gyr for G dwarfs, respectively; ii) the spin-down rate of F dwarfs is larger than that of G dwarfs; iii) stars with sub-solar metallicity tend to rotate faster and have the larger spin-down rate than the ones with super-solar metallicity at given mass bin. This suggests that the spin-down torque is affected by metallicity, i.e. opacity increases with metallicity, resulting in a deeper convective envelope and greater loss rate of angular momentum for metal-poor stars at the given masses \citep{Amard2020metal-spin}. In Fig.\ref{fig: PTD}(e), we notice that the solar $P_{rot}$ (24.47 $\sim$ 35 days) is larger than the median value of other stars, which is probably associated with the detection threshold of Kepler. It will be discussed in Section \ref{subsec: RT}. 

For K dwarfs in Fig.\ref{fig: PTD}(g-i), we find that: i) the correlation between age and $P_{rot}$ is not significant; ii) there are a fraction of fast rotators ($P_{rot} < $ 20 days) with high variability amplitudes ($logS_{ph}>3.3$), especially for K dwarfs, which may be affected by high activity level and large intrinsic stellar noise \citep{cdpp_2012PASP..124.1279C, Sph_P_knee_2022ApJ...933..195M}. iii) a notable dense region appears in $P_{rot}$ $\sim$ 25 days, indicative of low activity levels. This flat region in the longer period of K dwarfs may be a crucial prediction for the weakened magnetic braking \citep{VanSaders2016Natur.529..181V, pileup_2022ApJ...933..114D}.

\subsection{$R^{+}_{HK}$($\tau$, M, [Fe/H])} \label{subsec: RT}

The chromospheric activity levels decrease with increasing age, but the correlation of $\log R^{+}_{HK}-\tau$ differs due to mass and metallicity, as displayed in Fig.\ref{fig: RTD}. F dwarfs with sub-solar metallicity present a prominent negative correlation in the range of 2-6 Gyr as shown in Fig.\ref{fig: RTD}(a). Ones with [Fe/H] $>$ $-$0.1, the correlation of $\log R^{+}_{HK}-\tau$ is not significant, as illustrated in Fig.\ref{fig: RTD}(b-c). The metal-poor stars have shallower Ca II H\&K profiles that correspond to higher levels of $R^{+}_{HK}$ \citep{Rocha_rhk1_feh_1998MNRAS.298..332R, age-mass-metallicity-activity_2016A&A...594L...3L}.

For G dwarfs in Fig.\ref{fig: RTD}(d-f), there is a clear correlation between $\log R^{+}_{HK}$ and age within the range of 2-13 Gyr. As shown in Fig.\ref{fig: RTD}(e), the activity level of the Sun is comparable to that of G dwarfs with age near 4.5 Gyr, aligned with the previous findings \citep{Wright_2004ApJS_R'hk, Sreejith_R'hk_2020_AA,silva_R'hk_2021A&A...646A..77G}.  This means the Sun is a normal star at the stage, which likely contradicts the result of Fig.\ref{fig: PTD}(e). This suggests that it is difficult to detect the $P_{rot}$ of stars with solar activity level in Kepler data \citep{zhang_solar_type_2020ApJ...894L..11Z}. In addition, we select stars with masses ranging from 0.98 to 1.02 M$_{\odot}$ and metallicity in the range -0.05 to 0.05 dex, then give the slope of activity-age relation. We find our slope $\sim -0.53$ is similar to that $\sim -0.52$ from \cite{Lorenzo_2018A&A...619A..73L}.

In Fig.\ref{fig: RTD}(g-i), K dwarfs exhibit the most noticeable decay rate of $\log R^{+}_{HK}$ for the stars with [Fe/H] $>$ $-$0.1 than that of FG dwarfs in a range of 3-13 Gyr. The trend has been identified in open clusters as reported by \cite{zhangjiajun2019}, where later-type dwarfs exhibit a more pronounced trend of $\log R^{+}_{HK}-\tau$. This means that $\log R^{+}_{HK}$ is a powerful age probe for K dwarfs when the $P_{rot}-\tau$ fails. In addition, our results prove the feasibility of the chromospheric activity index as an age indicator for old field stars \citep{old_field_stars_2019ApJ...871...39M}.

Sub-solar metallicity F dwarfs exhibit an exceptionally larger decay rate of chromospheric activity than other stars with similar spectral types or metallicity. Since their surface convection zone is very thin, we need to consider the impact from the stars above the Kraft break \citep{Kraft_break_1967ApJ...150..551K}, therefore we further subdivide the stars in this interval into two mass bins (1.4-1.2, 1.2-1.1 M$_{\odot}$), which corresponding to above the Kraft break (Fig.\ref{fig: KbRTD}(a)) and below the Kraft break (Fig.\ref{fig: KbRTD}(b)). Fig.\ref{fig: KbRTD} reveals that the decay rate of $logR^{+}_{HK}$ is stronger for more massive or hotter stars, but the slope becomes smaller for stars below the Kraft break. The difference between the slopes confirms that the stars above the Kraft break (M $>$ 1.2 M$_{\odot}$) are probably associated with the strong negative correlation in Fig.\ref{fig: RTD}(a).

\section{Conclusion} \label{sec: summary}
To calibrate the relations of $P_{rot}$($\tau$, M, [Fe/H]) and $R^{+}_{HK}$($\tau$, M, [Fe/H]), we constructed a catalog of 52,321 FGK dwarfs, with $T_{\rm eff}$, [Fe/H], $M_G$, L/L$_{\odot}$, $[\alpha/Fe]$, $P_{rot}$, $S_{ph}$, $R^{+}_{HK}$, mass and age, etc. Our results demonstrate:

\begin{itemize}
    \item For F dwarfs, there is a clear relation among $P_{rot}$, age, and [Fe/H] in the range of 1-7 Gyr. However, the decrease of $\log R^{+}_{HK}$ over time is not significant except for those with [Fe/H] $<$ $-$0.1. 

    \item For G dwarfs, both the correlation of $P_{rot}-\tau$ and $\log R^{+}_{HK}-\tau$ are significant in the ranges 2-11 and 2-13 Gyr, respectively. This suggests that $P_{rot}$ and $\log R^{+}_{HK}$ are reliable age probes. 

    \item For K dwarfs, they behave the weakest correlation of $P_{rot}-\tau$. However, they exhibit the most prominent correlation of $\log R^{+}_{HK}-\tau$ than that of FG dwarfs within the age range of 3-13 Gyr.

    \item For the Sun, we confirm it has a comparable activity level with other G dwarfs at the given age.
\end{itemize}

We find that $P_{rot}$ is a reliable age indicator for FG dwarfs, while $\log R^{+}_{HK}$ is a powerful age probe for GK dwarfs, based on the relations of $P_{rot}$([Fe/H], M, $\tau$) and $R^{+}_{HK}$( [Fe/H], M, $\tau$). The calibrated relations expand the application of gyrochronology to 11 Gyr and magnetochronology to 13 Gyr. The results are promising in quick age estimation for the larger dwarf sample in forthcoming photometric and spectroscopic surveys.

This work is supported by the Joint Research Fund in Astronomy (U2031203) under cooperative agreement between the National Natural Science Foundation of China (NSFC) and Chinese Academy of Sciences (CAS), and NSFC grants (12090040, 12090042). J.-H.Z. acknowledges support from NSFC grant No.12103063. This work is partially supported by the Scholar Program of Beijing Academy of Science and Technology (DZ:BS202002) and the CSST project. M.X. acknowledges China National Key R\&D Program No.2022YFF0504200 and NSFC grant No.2022000083 for financial support for the computation of LAMOST spectroscopic stellar parameters with DD-Payne.

This work is based on data from the Guoshoujing Telescope (the Large Sky Area Multi-Object Fiber Spectroscopic Telescope LAMOST). Funding for the project has been provided by the National Development and Reform Commission. LAMOST is operated and managed by the National Astronomical Observatories, Chinese Academy of Sciences. This work has made use of data from the European Space Agency (ESA) mission Gaia (\url{https://www. cosmos.esa.int/gaia}), processed by the Gaia Data Processing and Analysis Consortium (DPAC, \url{https://www. cosmos.esa.int/web/gaia/dpac/consortium}). Funding for the DPAC has been provided by national institutions, in particular, the institutions participating in the Gaia Multilateral Agreement. This paper includes data collected by the Kepler mission. Funding for the Kepler mission was provided by the NASA Science Mission directorate. M.X. acknowledges China National Key R\&D Program No. 2022YFF0504200 and NSFC grant No. 2022000083 for financial support for the computation of LAMOST spectroscopic stellar parameters with DD-Payne.

\bibliography{sample631}{}
\bibliographystyle{aasjournal}
\end{document}